\documentclass[proceedings]{rmaa}

\title{SHALLOW CORES IN THE DARK MATTER HALOS: SELF-INTERACTION
IN ACTION?}

\author{Vladimir Avila-Reese\altaffilmark{1}, Claudio Firmani\altaffilmark{1,2},
 Elena D'Onghia\altaffilmark{3}, and Xavier Hern\'andez\altaffilmark{4}
       \affil{\it Submitted: June 5, 2000}}
\altaffiltext{1}{Instituto de Astronom\'{\i}a, Universidad Nacional 
        Aut\'onoma de M\'exico, M\'exico}
\altaffiltext{2}{Osservatorio Astronomico di Brera, Italy}
\altaffiltext{3}{Universita degli Studi di Milano, Italy}
\altaffiltext{4}{Osservatorio Astronomico di Arcetri, Italy}
\fulladdresses{
\item V. Avila-Reese and C. Firmani:
Instituto de Astronom\'{\i}a, UNAM, A.P. 70-268, 04510 M\'exico,
D.F., M\'exico 
\item E. D'Onghia: Universita degli Studi di Milano, via 
Celoria 16, 20100 Milano, Italy}

\shortauthor{Avila-Reese, Firmani, D'Onghia, \& Hern\'andez}
\shorttitle{Shallow cores in the dark matter halos}


\resumen{En base a datos observacionales para una muestra de galaxias
dominadas por materia obscura y un par de c\'umulos de galaxias, hemos
encontrado que la densidad central del halo no depende de su masa
y el radio del n\'ucleo es proporcional a la velocidad de rotaci\'on
m\'axima. Estas leyes de escala est\'an de acuerdo con 
halos CDM cuyas partes internas se expandieron por inestabilidades
gravot\'ermicas si las part\'{\i}culas autointeract\'uan eficazmente
s\'olo en estas partes. Encontramos que la secci\'on de choque es 
inversamente proporcional a la dispersi\'on de velocidades.}

\abstract{Using observational data for a sample of dark matter 
dominated galaxies and two cluster of galaxies, we have found that
the central halo density does not depend on its mass, and the core radius
is roughly proportional to the maximum rotation velocity. 
A good agreement with these scaling laws is obtained
for CDM halos whose dense inner parts were expanded by gravothermal
instabilities if the particles efficiently self-interact only in these
parts. We find that the particle cross-section is inversely proportional
to the velocity dispersion.}

\nonstopmode

\begin{document}
\maketitle

\section{Introduction}

During several decades, dynamical studies of galaxies, and group
and cluster of galaxies have pointed out to the existence of massive dark 
matter halos. 
On the other hand, according to current models of 
structure formation in the universe, luminous galaxies should form
from the gas trapped within the deep gravitational wells of 
dark matter (DM) structures emerged from primordial density fluctuations.  
In these models, non-dissipative, cold, collisionless particles 
(cold dark matter, CDM) were required. The CDM structure 
formation scenario succesfully 
accounted for a wide range of observations, in particular on large scales.
However, on small scales, compared with observations this scenario
seems to predict too centrally concentrated halos and too 
much substructure in Milky Way-size halos. These discrepancies have 
induced to introduce some modifications to the CDM scenario, in 
particular, to the nature of the DM. 

From the point of view of particle physics, a large list of candidate
DM particles has been proposed but unfortunately, none of
the particles that might constitute the universe's missing mass have
been detected nowadays. Nevertheless, it is possible that astronomical
observations may help us to constrict some of the properties of these 
particles. For example, as was mentioned above, the existence of 
soft cores in the dark halos appears to be not compatible with collisionless
CDM particles. Therefore, astronomical studies about the halo properties
--- in particular of their cores --- 
are crucial for understanding the nature of the dark particles and
the structure formation in the universe, as was emphasized in a 
pioneering paper on this subject by J. Kormendy (1990; see also
Kormendy 1988). Here,  we
summarize the halo core scaling relationships we have inferred from 
observations from dwarf galaxies through cluster of galaxies (Firmani
et al. 2000a,b), and
we dicuss some of the implications of our results on the nature
of the DM particles and the formation of halos.

\section{Halo core scaling laws from observations}

Analysis of the original virialized {\it halo mass distribution} for most of 
galaxies is uncertain due to the ambiguities in the estimate of the 
stellar mass-to-light ratios $M/L$ and the gravitational pull the
collapsing gas exerted over the inner parts of the halo.
This is why we limited our study to only galaxies (i) strongly dominated
by DM and (ii) with accurately measured rotation curves. The sample 
taken from the literature consists 
of six dwarf galaxies, nine LSB galaxies, and two late-type low luminosity 
galaxies (Firmani et al. 2000b). In all these cases, the galaxies are 
DM dominated. Even so, we have substracted from the observed rotation 
curve the small disk 
contribution. The halo components were fitted to a non-singular isothemal
model; thus, for each galaxy characterized by its maximum circular
velocity V$_{\rm max}$\, we estimate its central density $\rho_c$ 
and core radius $r_c$.  For the less DM 
dominated LSB and low-luminosity galaxies, we have 
roughly calculated the factor by which the halo component was ``deformed'' 
due to the disk pull over the DM using the adiabatic invariance 
approximation (see details in Firmani et al. 2000b).
On galaxy cluster scale, we have used the surface mass distribution for the 
cluster CL0024+1654 derived from unprecedent high-resolution strong 
lensing mass maps (Tyson, Kochanski, \& Dell'Antonio 1998), and for the cluster
CL0016+16 derived from weak lensing studies (Smail et al. 1995). In both
cases there is no evidence of a massive cD galaxy and the inner mass
distribution is soft.

In Figs. 1 we show the dependence of $r_c$ on 
V$_{\rm max}$ we have found from the observational data. Although 
with a big scatter, within a large range in V$_{\rm max}$ we estimate that:
\begin{equation}
\rho_c (r)\approx 0.02 \ {\rm M_{\odot}pc^{-3}}\ \ \ \ \ \ \
{\rm and}\ \ \ \ \ \ \ r_c\approx 
5.5\Bigl[\frac{V_{\rm max}}{100 {\rm kms^{-1}}}\Bigr]^{0.95} \ {\rm kpc}.
\end{equation}
Similar results were found for an uniform sample of high and LSB 
galaxies of the Coma Ursa Mayor cluster (Verheijen 1997; \S 6). In 
this case, the rotation curve decompositions were made assuming 
$M/L_K$ constant for all galaxies, and the halo component was
fitted to a pseudo-isothermal model.  In contrast, from a sample
of Sc-Im and dwarf galaxies, Kormendy (1988,1990) inferred that 
$\rho_c$ decreases with the galaxy luminosity (or V$_{\rm max}$). Certainly, more
efforts should be done in the future in order to increase the sample
of objects and to reduce the uncertainties in the rotation
curve decomposition techniques. We remark the importance of strong
gravitational lensing studies in order to directly probe the inner
regions of the cluster of galaxies.

\section{Implications of the inferred halo core scaling laws}

The existence of soft halo cores and even more, the scaling
laws obtained for DM dominated systems [eq.(1)], are in complete
disagreement with the predictions of CDM models. Warm dark matter
(WDM) has been proposed in order to solve the other conflict of the
CDM scenario ---the overlying number of guest (satellite) halos
in a Milky Way-size halo. Cosmological N-body simulations have 
shown that the latter problem is indeed solved for a filtering
scale in the power spectrum of $\sim 0.1$ Mpc which corresponds
to a warm particle of $\sim 1$ KeV (Col\'{\i}n, Avila-Reese,
\& Valenzuela 2000). These authors have also shown that the
density profiles of halos with masses much larger than that 
corresponding to the filtering scale ($\sim 10^9$ h$^{-1}$M$_{\odot}$)
are very similar to those of the CDM models (see also Moore et al.
1999). Thus, even if the halos with masses near or smaller than the
filtering mass would have a core, the more massive WDM halos will 
not obey the scaling laws inferred from observations in $\S 2$
(see also Avila-Reese, Firmani, \& Hern\'andez 1998). 

\begin{figure}

\vspace{7.8cm}
\includegraphics{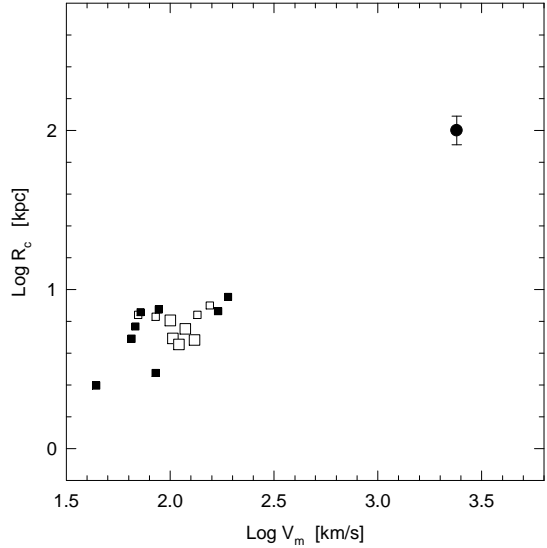}
    \caption{The core radius $R_c$ vs. V$_{\rm max,100}$ for dwarfs
(filled squares), LSB galaxies (empty squares) and the CL0024+1654 cluster
(filled circle).}
\end{figure}

 Spergel \& Steinhardt (2000) suggested other modification to 
the nature of the DM particles: the introduction of self-interaction. 
In Firmani et al. 2000a, the {\it gravothermal expansion} was 
proposed as the mechanism able to produce soft cores in self-interacting 
CDM halos. The inner  velocity dispersion profile of these halos 
raises with radius.  Therefore, if 
particles are collisional, heat transfers inwards, the core 
expands and cools, exacerbating even more the temperature gradient.
This process is similar to the postcollapse gravothermal oscillations
in globular clusters (Bettwieser \& Sugimoto 1984; Goodman 1987).
For globular clusters, the core expansion halts when the inner
dispersion velocity profile flattens; this occurs because 
there is also an outwards heat flux from the maximum of the velocity profile. 
In the case of DM halos, we propose a collisional cross-section
$\sigma $ such that self-interaction is efficient only in the more 
dense halo regions.
Besides, as the soft core grows, the core density decreases and
at some moment, self-interaction should become inefficient even in 
the inner regions. On the other hand, it is important
to beer in mind that the CDM halo does not form instantaneously,
but by a hierarchical mass aggregation process which establishes a
cuspy inner structure with a positive velocity dispersion gradient. 

Recent numerical simulations for a halo with Hernquist density
profile and with relatively small cross-sections per unit of the 
particle mass $m_X$ ($\sigma _{\star} =\sigma /m_X$) have shown that the 
gravothermal processes act in time scales that depend on the value
of $\sigma _{\star}$ (Burkert 2000; Kochanek \& White 2000; see also Quinlan 1996).
An important constriction is that the halo lifetime should be in between
the core expansion time and the core collapse time; otherwise either
the shallow core still have not been formed or the core is already
in its collapse phase. In Firmani et al. (2000a,b), using the average
observed $\rho_c$ and supossing that the collision time $t_{\rm col}$ in 
the core is close to the Hubble time, a lower limit for 
$\sigma _{\star}$ was estimated\footnote{Here we assume that 
V$_{\rm max}\approx v_{\rm rms,max}$; in fact, for CDM halos
V$_{\rm max}$ is roughly 1.3-1.7 times larger than $v_{\rm rms,max}$}:
$\sigma _{\star} \approx 4 \ 10^{-25}V_{\rm max,100}^{-1}$  cm$^2$/GeV, 
where V$_{\rm max,100}$ is V$_{\rm max}$ in units of 100 km/s. An important
point to be noted is that  $\sigma _{\star}$ depends on V$_{\rm max}$ or the
maximum velocity dispersion, i.e. the cross-section is a function of 
the particle energy as in other classical physical interactions.
For velocity dispersions corresponding to galaxy clusters, this value is
close to the limit estimated by Miralda-Escud\'e (2000) from the
observationally inferred ellipticity of the cluster MS2137-23.

One may think that the evolution of the collisionless DM halo occurs
in scales of dynamical times, $t_{\rm dyn}$, while those central regions 
of the halo affected by the gravothermal processes, evolve in 
relaxation time scales, $t_{\rm rel}$. The final halo density profile 
is the result of both dynamical processes. The simulations
carried out by Burkert (2000) and Kochanek \& White (2000; KW00) are for
a halo already virialized. Therefore, these simulations do not
describe the cosmological process of halo collapse and virialization. 
Kochanek \& White find that the gravothermal core collapse occurs
in scale times less than $\sim 5$ times the core formation time $t_c$
independent of the value of $\sigma _{\star}$. On the other hand,
$t_c\propto 1/\sigma _{\star}$.
Thus, for $\sigma _{\star}$ small enough the halos may still be
in their core expansion phase. Besides, if $\sigma _{\star}$
depends on V$_{\rm max}$ as we have inferred 
from observations, then larger halos should be today in earlier 
stages of gravothermal expansion than smaller halos, i.e. their 
central densities have not decreased too much. This, combined 
with the fact that in the hierarchical scenario smaller halos are 
intrinsically more concentrated than larger ones, could produce
the invariance of $\rho _c$ with the halo scale.   

The simulation of KW00 are for a Hernquist halo,
and they express $\sigma _{\star}$ in unities of $r_H^2/M_h$, where
$M_h$ is the halo mass and $r_H$ is the scale radius of the Hernquist
profile. Fitting the Hernquist profile to halos obtained in a 
N-body CDM simulation, one finds that $r_H^2/M_h$ is roughly constant. 
In order to obtain more quantitative estimates, we have used results
for a $\Lambda$CDM$_{0.3}$, h=0.7 model (Avila-Reese et al. 
1999). We calculate $r_H$ as
$r_H=r_v/c_H$, where the virial radius $r_v$ is defined as the radius 
where the average halo density is $\Delta _c$ times the background
density (for our cosmology, $\Delta _c=340$), and $c_H$ is the 
concentration parameter which ultimately depends on the halo mass or 
V$_{\rm max}$ and is the only free parameter in the cosmological halo density
profiles. From the results of the simulation, we find on the average 
c$_H=37.5/($V$_{\rm max}/{\rm kms^{-1}})^{0.36}$. The virial 
radius is proportional to V$_v$, the circular velocity at this 
radius, and for the Hernquist profile, V$_v=2$V$_{\rm max}$c$_H^{1/2}/(1+c_H)$. 
We find that 
$r_H^2/M_h\approx 7 \ 10^{-24}$ cm$^2$/GeV. Thus, in KW00
$\sigma _{\star}$ would be $\hat{\sigma}\times 7 \ 10^{-24}$ cm$^2$/GeV. 
For $\hat{\sigma}=1$, KW00 find that $t_c\sim 1.7$ dynamical times;
after this the halo suffers the gravothermal core collapse. For a value
of $\sigma _{\star}$ as that we have inferred from observations, for a 
V$_{\rm max}\approx 100$ km/s halo for example ($\sigma _{\star}\approx
4 \ 10^{-25}$ cm$^2$/GeV), $\hat{\sigma}\sim 0.05$. The dynamical 
time (as deffined in KW00) for a V$_{\rm max}\approx 100$ km/s halo is $\sim
5 \ 10^8$ years. Therefore, the core formation time would be
of the order of a Hubble time. Halos larger than V$_{\rm max}\approx 100$ km/s
would have even larger core formation times. 

\section{Conclusions}

$\bullet$ The halo core scaling laws inferred from observations of
dwarf galaxies to galaxy clusters show 
that $\rho _c$ does not depend on the halo mass or V$_{\rm max}$
and the core radius is roughly proportional to V$_{\rm max}$.

$\bullet$ If the dark particles are self-interacting with not very 
large cross sections, then gravothermal processes may produce
a soft core in the DM halos. Using the observational data,
we estimated the value of $\sigma _{\star}$ and found that is 
roughly proportional to 
V$_{\rm max}^{-1}\propto v_{\rm rms,max}^{-1}$.

$\bullet$ Results from numerical simulations of already virialized halos
with self-interaction, show that if $\sigma _{\star}$ is
of the order we inferred from observations, then $t_c$ for 
small halos is close to the Hubble time, while for larger
halos, $t_c$ is probably even larger, i.e. these halos
are still in early stages of gravothermal expansion. 
Numerical simulations and theoretical studies of collapsing
and virializing DM halos where self-interacion is efficient 
only in the more dense inner regions are necessary in order
to attain more quantitative conclusions.

\end{document}